\documentclass[%
 reprint,
 amsmath,amssymb,
 aps,
]{revtex4-1}

\usepackage{graphicx}% Include figure files
\usepackage{dcolumn}% Align table columns on decimal point
\usepackage{bm}% bold math
\usepackage{textcomp}
\usepackage{units}
\begin{document}

\preprint{APS/123-QED}

\title{Coherent properties of a tunable low energy electron matter wave source}

\author{A. Pooch$^1$, M. Seidling$^1$, N. Kerker$^1$, R. R\"{o}pke$^1$, A. Rembold$^1$, W.T. Chang$^2$, I.S. Hwang$^2$ and A. Stibor$^1$}
\affiliation{$^1$Institute of Physics and Center for Collective Quantum Phenomena in LISA$^+$,
University of T\"{u}bingen, Auf der Morgenstelle 15, 72076 T\"{u}bingen, Germany\\
$^2$Institute of Physics, Academia Sinica, Nankang, Taipei 11529, Taiwan, Republic of China}

\date{\today}% It is always \today, today,

\begin{abstract}
A general challenge in various quantum experiments and applications is to develop suitable sources for coherent particles. In particular, recent progress in microscopy, interferometry, metrology, decoherence measurements and chip based applications rely on intensive, tunable, coherent sources for free low energy electron matter waves. In most cases, the electrons get field emitted from a metal nanotip where its radius and geometry towards a counter electrode determines the field distribution and the emission voltage. A higher emission is often connected to faster electrons with smaller de Broglie wavelengths, requiring larger pattern magnification after matter wave diffraction or interferometry. This can be prevented with a well-known setup consisting of two counter electrodes that allow independent setting of the beam intensity and velocity. However, it needs to be tested if the coherent properties of such a source are preserved after the acceleration and deceleration of the electrons. Here, we study the coherence of the beam in a biprism interferometer with a single atom tip electron field emitter if the particle velocity and wavelength varies after emission. With a Wien filter measurement and a contrast correlation analysis we demonstrate that the intensity of the source at a certain particle wavelength can be enhanced up to a factor of 33 without changing the transverse and longitudinal coherence of the electron beam. In addition, the energy width of the single atom tip emitter was measured to be \unit[377]{meV}, corresponding to a longitudinal coherence length of \unit[82]{nm}. The design has potential applications in interferometry, microscopy and sensor technology.
\end{abstract}

\maketitle

\section{Introduction}

The quest to find the optimal particle source for a specific application in quantum physics has determined the progress in matter wave experiments for atoms \cite{Cronin2007,Anglin2002}, neutrons \cite{Rauch2015}, molecules \cite{Hornberger2012,Juffmann2013}, electrons \cite{Hasselbach2010,Batelaan2007} and ions \cite{Hasselbach2010,Hasselbach1999}. New sources for free electron waves led to sophisticated recent developments in microscopy \cite{Putnam2009,Kruit2016,Chang2015,Latychevskaia2017,Longchamp2015}, laser induced single particle interference \cite{Ehberger2015} or time-resolved dephasing measurements \cite{Rembold2014,Guenther2015,Rembold2017}. Particularly, to study the Coulomb-induced decoherence of an electron superposition close to a metallic, semi- or superconducting surface \cite{Sonnentag2007,Machnikowski2006,Scheel2012,Anglin1997}, a source is desirable that allows the tuning of the electrons energy by remaining their coherence. The energy determines the velocity and therefore the interaction time of the quantum state with the environment. Most of the sources in interference experiments so far were etched metal tips with a diameter of several ten nanometers. By adding a monolayer of iridium or palladium the emission area can be reduced down to the size of a single atom at the end of a pyramidal atom stack (single atom tip (SAT) \cite{Kuo2006,Kuo2008,Schuetz2014}). The electron field emission follows the theory of Fowler and Nordheim \cite{Fowler1928,Kuo2006,Schuetz2014} and the extraction voltage is determined by the properties of the tip, such as the material and the geometry, in relation to the distance and dimensions of a counter electrode. It is difficult to fabricate the geometry of a tip exactly on the nanometer scale, even if progress was made to control the tip profile during electrochemical etching \cite{Chang2012}. As a result, individual tips have different extraction voltages for which the field emission will start, leading to varying intensity to velocity relations between different tips. 

For that reason, in electron biprism interferometers the tip radius was manufactured as small as possible to get a low extraction voltage resulting in an intense and spatially coherent beam with a large electron wavelength. The electron emission signal was then enhanced by increasing the tip voltage. However, as a consequence, the resulting matter waves have larger energies and therefore shorter wavelengths, leading to smaller diffraction or interference fringes. Assuming a limited detection resolution and area, this in turn requires larger pattern magnification which again reduces the signal. This problem can easily be addressed with a well-known technique used in electron microscopes by implementing two counter electrodes (apertures) behind the tip. They allow to control the velocity (and wavelength) of the electrons independently to the emission intensity. By setting a low tip voltage in combination with a high first counter electrode voltage relative to a second grounded aperture, a high emission intensity of slow electrons can be realized. However, this method was not applied in most biprism interferometers so far. The reason could be that it remains unclear, how the lateral and longitudinal coherence of the particles are affected by accelerating and decelerating the particle, since the wavelength, defining the coherence of the source, is not constant any more after emission. Therefore, in this geometry a combination of two factors influence the electrons. First, there is a position dependent change of the electrons velocity leading to different wavelengths. And second, there is a lens effect of the electrodes. 

In this article we describe such a beam source and test in a biprism electron interferometer how the lateral and longitudinal coherence of the particles are affected after tuning the intensity of the beam and keeping the wavelength constant after passing the second aperture. We demonstrate with an iridium SAT that neither lens effects of the apertures nor the acceleration and deceleration of the electrons change their lateral coherence properties. Furthermore, the longitudinal coherence length and the energy spread of the emitted electrons are measured with a Wien filter \cite{Nicklaus1993}. These properties also remain unchanged. Our source allows to increase the coherent low energy electron emission by a factor of 33 at a constant matter wavelength, resulting in a constant interference contrast, pattern periodicity and amount of fringes. The results have applications in matter wave interferometry \cite{Hasselbach2010,Schuetz2014}, in Aharonov-Bohm studies \cite{Aharonov1959,Schuetz2015b}, decoherence measurements \cite{Sonnentag2007,Machnikowski2006,Scheel2012,Anglin1997}, electron diffraction microscopy \cite{Chang2015,Latychevskaia2017,Longchamp2015}, time resolved ultrafast electron diffraction \cite{Gulde2014} and for the development of a quantum electron microscope \cite{Putnam2009,Kruit2016}. 

\section{Experiment}

\begin{figure}[t]
\centering
\includegraphics[width=0.5\textwidth]{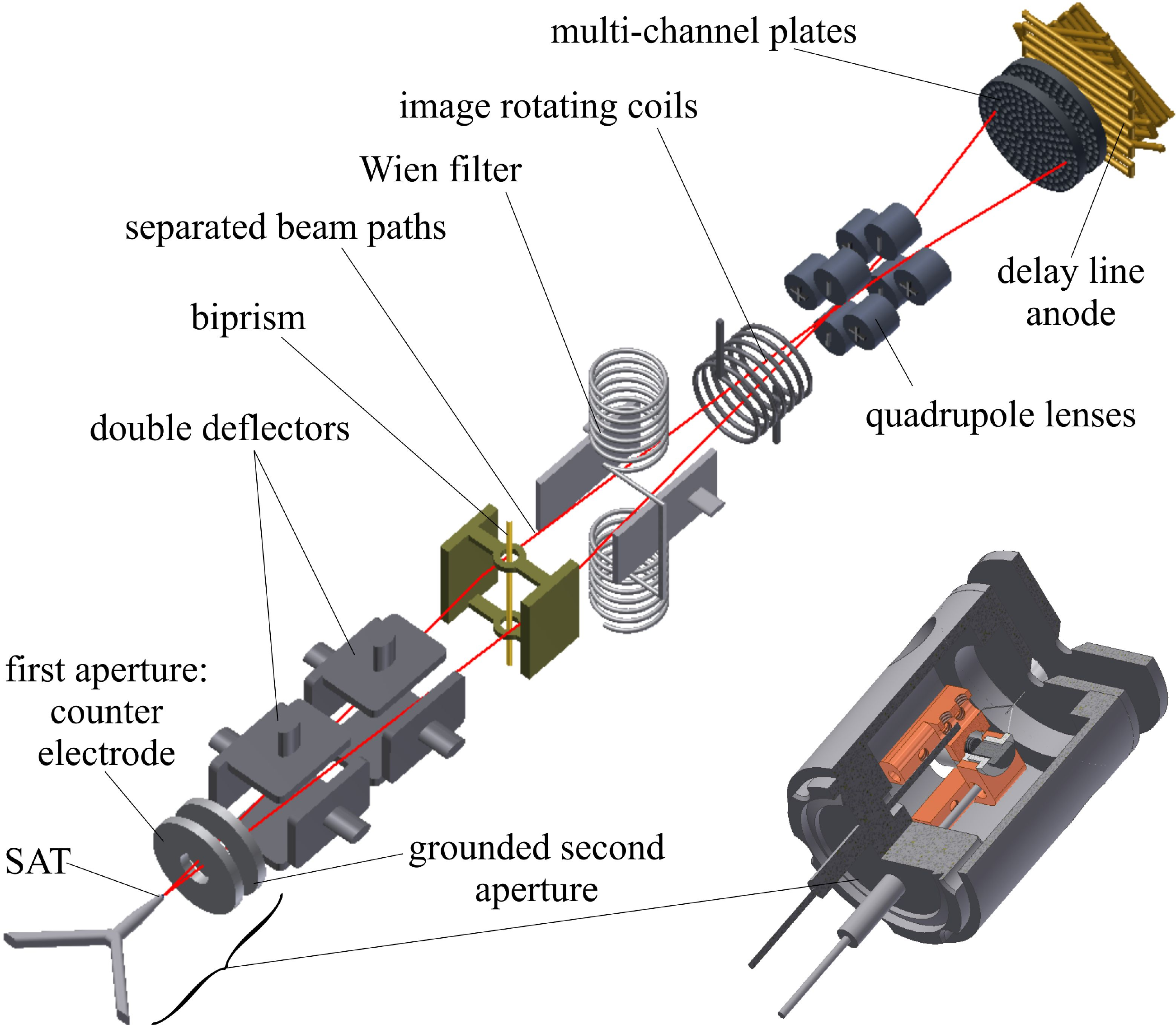} \caption{Sketch of the experimental setup and the separated beam paths (red line) to measure the coherent properties of the single atom tip at different tip and counter electrode settings and beam intensities in a biprism electron interferometer (not to scale). Inset: Mechanical drawing of the SAT field emission source realization in our setup with the two apertures (to scale).}
\label{fig1}
\end{figure}

A sketch of the experimental setup is illustrated in fig.~\ref{fig1}. The aim of the experiment is to increase the electron emission without changing the transversal and longitudinal coherence of the matter waves. The de Broglie wavelength, which is reciprocal to the particle velocity, should thereby be kept constant. This is possible with a configuration where the field emission tip, in our case a SAT \cite{Kuo2006,Kuo2008,Schuetz2014}, is set on a low negative potential $U_{SAT}$ and a counter electrode is set on a positive potential $U_c$. Since only the field between these components is relevant for the emission process, the electron beam intensity can be increased by raising the positive voltage of the counter electrode. In the following beam line, the electrons get decelerated by a second, grounded electrode to the low velocity corresponding to the electrical potential of the tip. Thereby, lower electron energies can be realized compared to the conventional high voltage field emission directly to a grounded aperture. This is a well-known setup. However, it needs to be demonstrated that the longitudinal and transversal coherence of the beam is not reduced by the accelerating and decelerating process.

To study the coherent properties of this emitter setup, it is integrated in a biprism electron interferometer. It includes several beam optic parts from a former experiment by Sonnentag et al.~\cite{Sonnentag2007}. The beam is adjusted by two double deflectors towards an electrostatic biprism. It consists out of a gold-palladium coated glass fiber with a diameter of \unit[400]{nm} \cite{Schuetz2014} between two grounded electrodes. Just like the optical biprism for light, the electrostatic biprism separates coherently the electron waves from the emitter \cite{Moellenstedt1956,Hasselbach2010}. By applying a positive voltage $U_f$ on the fiber, the biprism bends all beam paths by the same angle and combines them at the entrance of a magnifying quadrupole lens. The partial beams interfere and form a fringe pattern parallel to the biprism fiber. Directly after the biprism, the beam traverses a Wien filter, consisting of two opposing electrodes and two magnetic coils. As a result of the finite energy spread of the emitted electrons, the separated partial waves can be described by matter wave packages. The Wien filter allows to shift them longitudinally relative to each other and to measure thereby the longitudinal coherence length of the beam \cite{Nicklaus1993}. To align the orientation of the fringes towards the magnifying axis of the quadrupole lens, an image rotating coil is positioned around the beam path behind the Wien filter. The interference pattern is magnified by two quadrupole lenses in the transverse direction normal to the biprism fiber with a magnification factor of several thousand. The electrons are amplified by two multi-channel plates and detected with a delay line anode \cite{Jagutzki2002}. The position and point in time for every single electron event is recorded allowing a second-order correlation data analysis \cite{Rembold2017,Guenther2015,Rembold2014,Rembold2017b}. The red line in fig.~\ref{fig1} illustrates a possible beam path according to the wave-particle duality. The whole setup is in an ultrahigh vacuum chamber at a pressure of $<\unit[5\cdot 10^{-10}]{mbar}$ and magnetically shielded by a mu-metal tube.

\section{Simulation}

\begin{figure}[ht]
\centering
\includegraphics[width=0.5\textwidth]{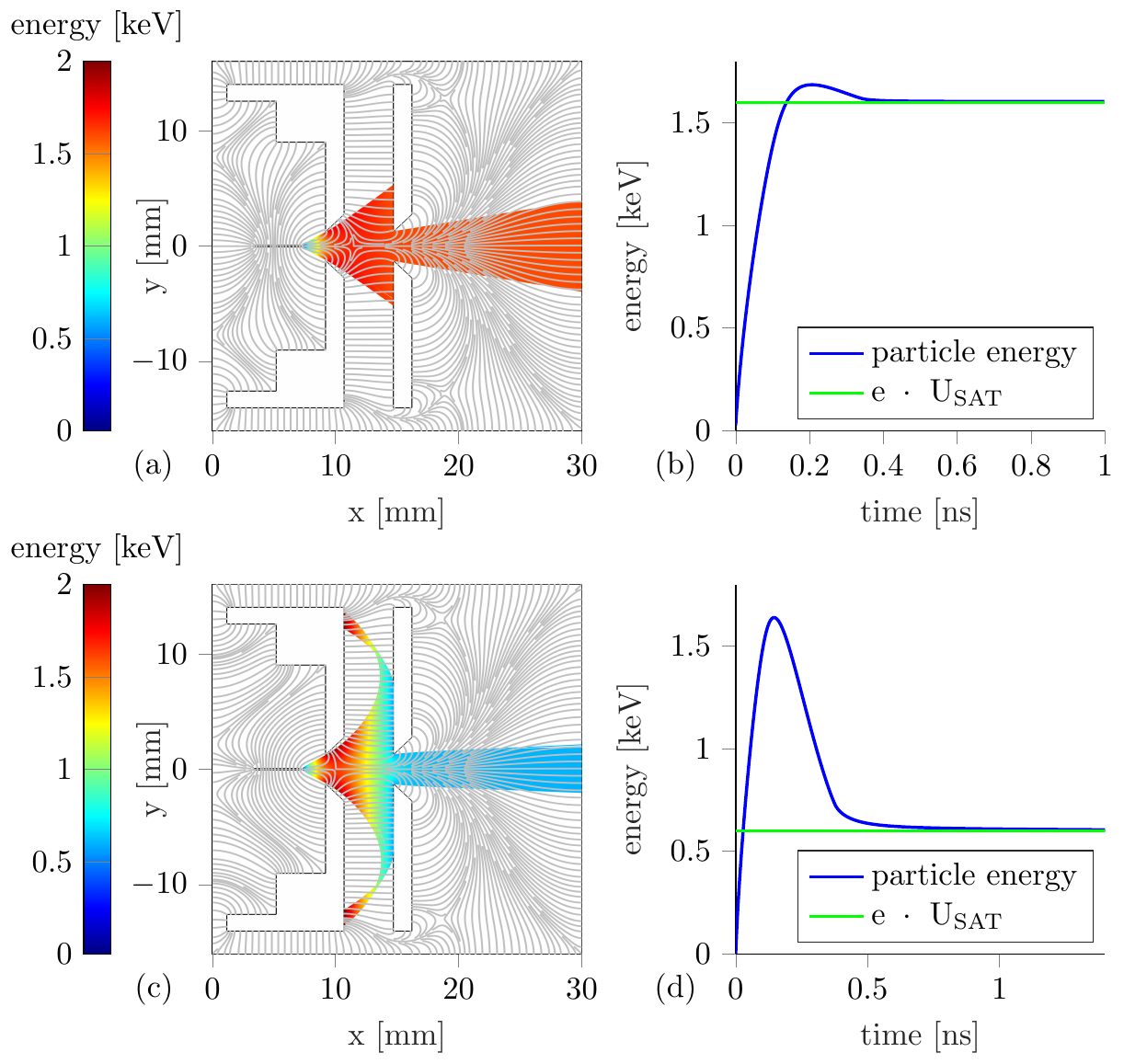} \caption{(a) Simulation of the electrical field distribution with a cross section of the cathode box geometry including the tungsten tip and the two apertures. The tip defines the optical axis at $y=0$. It is set on a potential of $U_{SAT}=\unit[-1600]{V}$ and emits electrons with \unit[0]{eV} starting energy. They accelerate in the electric field between the tip and the counter electrode which is set on a voltage $U_c=\unit[200]{V}$. The second aperture is grounded. The color bar represents the energy of the electrons. Also the electric field lines are shown in pale grey schematically. (b) The average energy of the electrons plotted versus time as they travel along the optical axis. They get accelerated by the first electrode and decelerated by the second, approaching the energy corresponding to the tip potential $e\cdot U_{SAT}=\unit[1600]{eV}$. (c) Same geometry as in (a) with the settings $U_{SAT}=\unit[-600]{V}$ and $U_c=\unit[1378]{V}$. Some of the electron trajectories are bend back onto the surface of the counter electrode. However, those electrons passing the second electrode end up at a significantly lower velocity compared to (a). (d) The acceleration and deceleration between the counter and second aperture is more distinct than in (b) leading to an electron energy of \unit[600]{eV}.}
\label{fig4}
\end{figure}

For a detailed description of the electrical field distribution between the tip, the counter and second electrode, simulations with the program \textit{Comsol} \cite{Comsol} were performed. The results are illustrated in fig.~\ref{fig4}. Two cases were studied: first, the tip is set on a potential of \unit[-1600]{V}, the counter electrode on \unit[200]{V} and the second aperture is grounded. This combination of voltages is used in most of our presented data. As can be seen in the cross section of fig.~\ref{fig4} (a) a $\unit[125]{\mu m}$ thick tungsten wire is orientated horizontally at $y=0$. Electrons with zero starting energy accelerate in $x$-direction. As it is simulated in fig.~\ref{fig4} (b), those electrons passing the two apertures with \unit[2.5]{mm} diameter keep moving on the optical axis and approach an energy of \unit[1600]{eV} given by the tip potential. Due to computational reasons the tip radius in this geometry is set to be $\unit[2]{\mu m}$, even if the actual physical tip radii are typically around \unit[50]{nm}. However, this does not change the results for the electric field lines further away from the tip. For the calculation of the electron trajectories the starting energy and direction of the emission is selected manually. The field emission itself could not be simulated and thus the exact field strength at the tip surface is not relevant. Certainly, it increases with increasing counter electrode voltage $U_c$. Fig.~\ref{fig4} also indicates that the coherent signal enhancement is not due to a lensing effect of the electron beam.

In the second case, simulated in fig.~\ref{fig4} (c), the tip is set on a low voltage of \unit[-600]{V}. To keep the relative potential high, \unit[1378]{V} were applied on the counter electrode. This setting emerged to be the limit for creating interference fringes with minimal tip voltage in the experiment. In the simulations it leads to particle trajectories where electrons even get deflected back onto the counter electrode. The electrons close to the optical axis reach the final energy of only \unit[600]{eV} in the interferometer with a corresponding large matter wavelength. As it will be verified in the next section, those slow electrons do not lose their coherent properties.

\begin{figure*}[ht]
\centering
\includegraphics[width=1.0\textwidth]{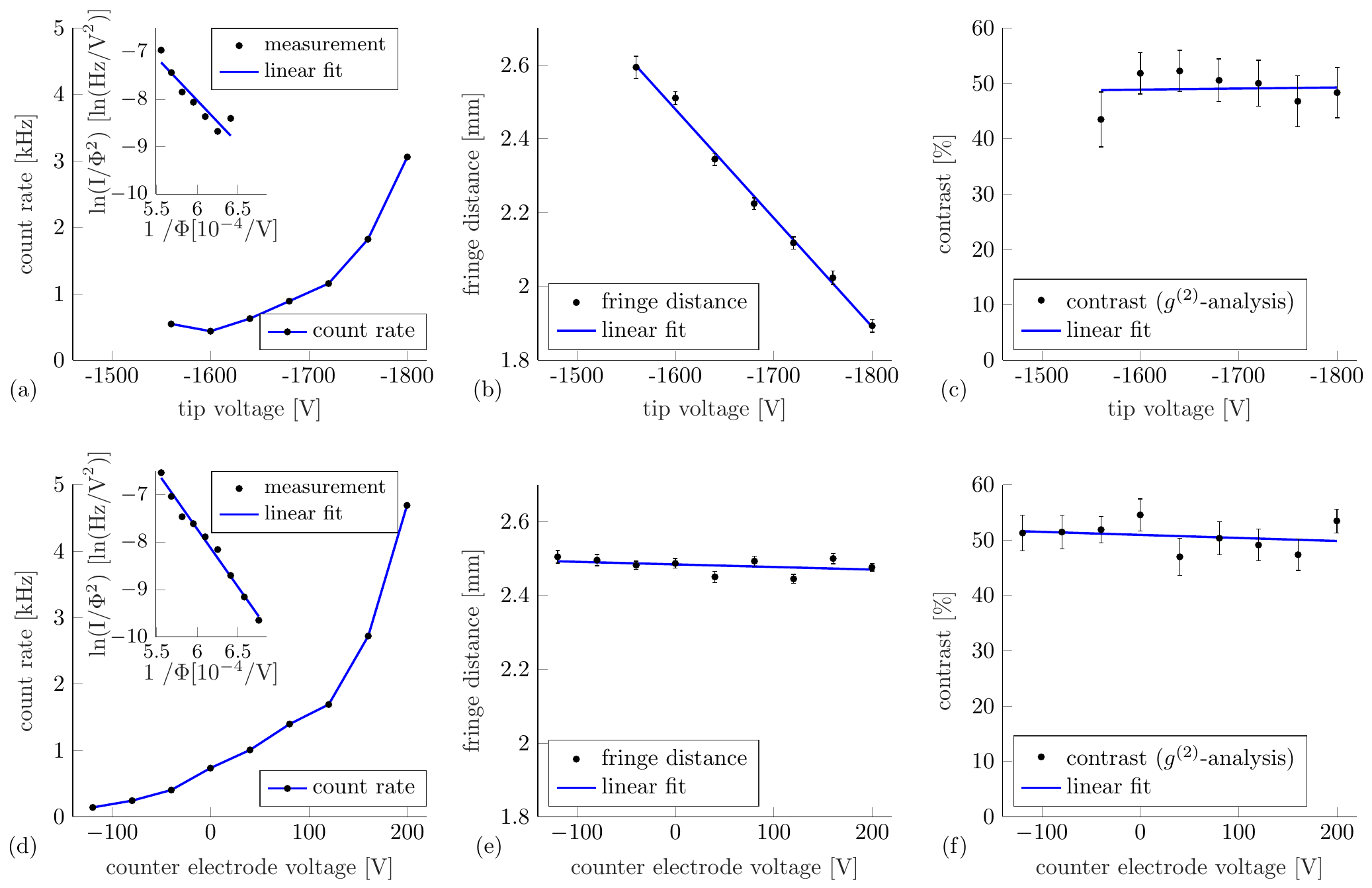} \caption{Comparison of the behavior in count rate, fringe distance and contrast between the increase of tip voltage towards a grounded counter electrode ((a) to (c)) and the application of a counter voltage to the first aperture ((d) to (f)). (a) Count rate at the detector plane versus the tip field emission voltage. Inset: plot according to the Fowler-Nordheim relation, where $\Phi$ equals the tip voltage. (b) Decrease of the measured fringe distances due to the higher electron energy, resulting in shorter de Broglie wave lengths and a lower quadrupole magnification. (c) The determined dephasing corrected contrast remains constant in this range of tip voltages. (d) In contrast to (a) the count rate is determined as a function of the aperture voltage, keeping the tip voltage constant at \unit[-1600]{V}. Inset: plot according to the Fowler-Nordheim relation, where $\Phi$ equals the potential difference between the tip and the counter electrode. (e) The fringe distance does not change over all different aperture voltages due to the constant electron energy according to the simulations in fig.~\ref{fig4} (b). (f) Also the interference contrast remains constant in the measurement region.
}
\label{fig2}
\end{figure*}

\begin{figure}[t]
\centering
\includegraphics[width=0.5\textwidth]{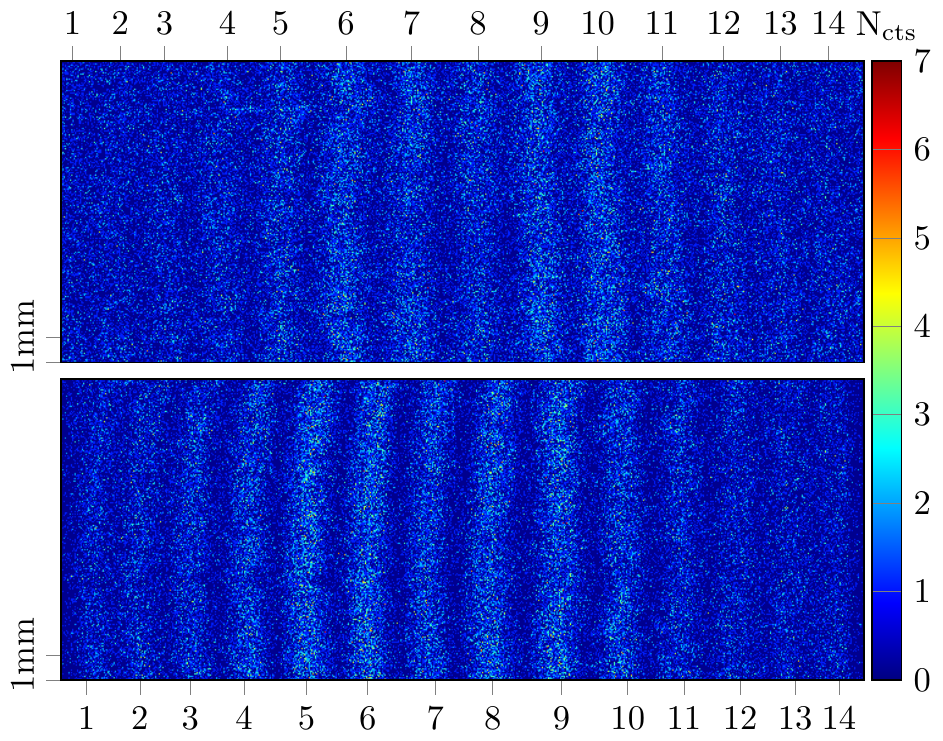} \caption{Comparison of two interferograms with different voltages at the counter electrode $U_c = \unit[-119.7]{V}$ (top picture) and $U_c = \unit[+199.7]{V}$ (bottom picture), both at a fix tip voltage of $U_{SAT}=\unit[-1600]{V}$. The interferograms exhibit the same amount of interference fringes (denoted by the number on the horizontal axis) and width of interference, revealing the transversal coherence of the interfering electrons is not influenced by the counter electrode voltage. A length scale is shown in the left bottom corner and the color bar represents the number of hits per pixel. Both pictures contain $3\times 10^5$ counts and are dephasing corrected by second-order correlation analysis \cite{Rembold2014,Guenther2015,Rembold2017,Rembold2017b}. The undisturbed interference contrasts were thereby determined to be \unit[51.3 $\pm$ 3.2]{\%} at \unit[-119.7]{V} and  \unit[53.5 $\pm$ 2.2]{\%} at \unit[+199.7]{V}. The original uncorrected contrasts in the spatial pattern, determined by the model function described in the text, were \unit[30.6 $\pm$ 2.1]{\%} and \unit[33.9 $\pm$ 2.2]{\%}, respectively.}
\label{figIntpat}
\end{figure}

\section{Results}
For the characterization of the intensity enhancement and the coherent properties of our field emission setup, we compared two situations. In the first one, the counter electrode is grounded and the tip voltage is increased. This is the usual operating mode in most biprism matter wave interferometers so far \cite{Hasselbach2010}. The second case demonstrates our method for the coherent signal enhancement. Thereby, the tip is set on a fixed potential and the first counter electrode is varied such as simulated in the last section. In both cases the second electrode is grounded. 

\begin{figure}[t]
\centering
\includegraphics[width=0.5\textwidth]{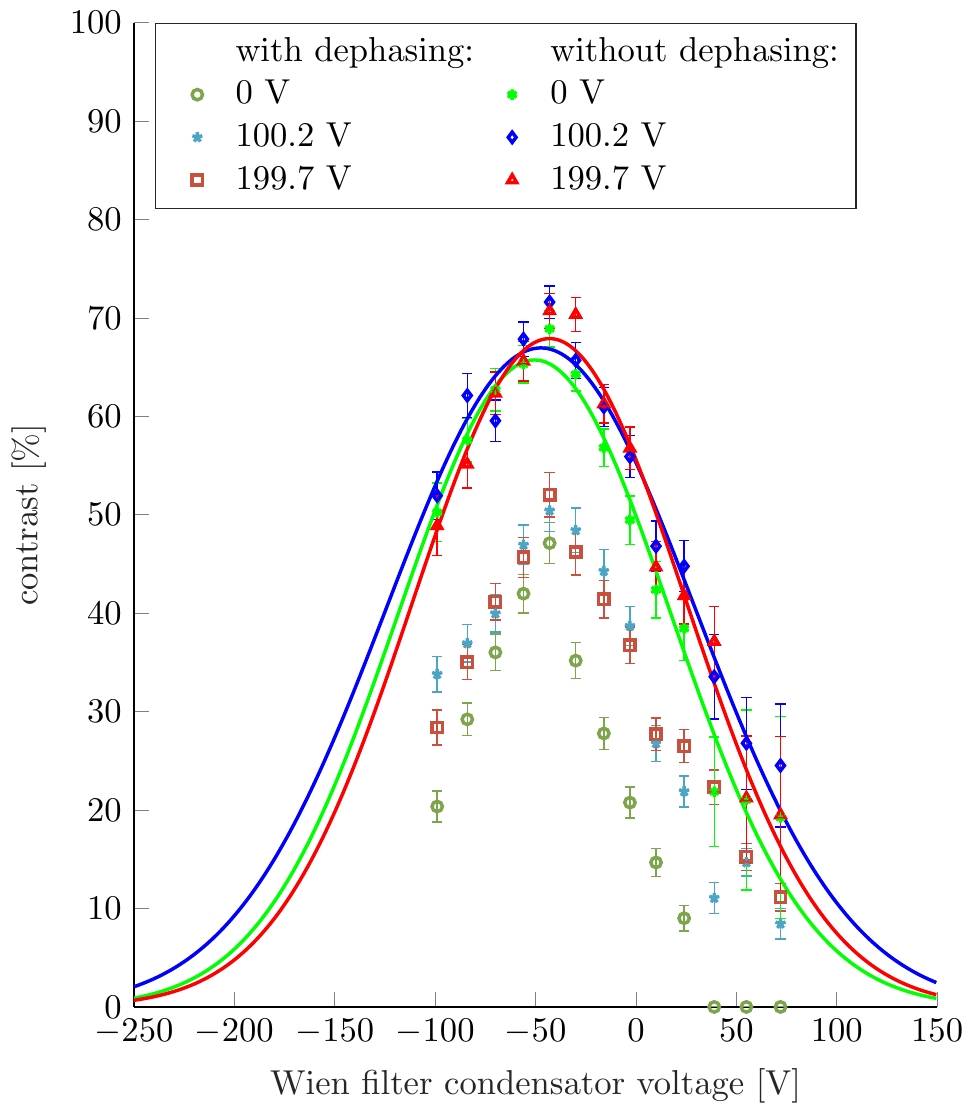} \caption{Measurement of the interference contrast as a function of different Wien filter condensator voltages for a tip potential $U_{SAT} = \unit[-1600]{V}$ and three different voltages for the first aperture ($U_c = \unit[0]{V}$, \unit[100.2]{V} and \unit[199.7]{V}). The determined contrast revealed by intensity evaluation of the spatial interference pattern is shown in green circles, blue stars and red squares, respectively. The data is strongly dephased by the network frequency at \unit[50]{Hz}. For that reason a $g^{(2)}$-correlation analysis using spatial and temporal differences of particle events at the detector \cite{Rembold2014,Guenther2015,Rembold2017,Rembold2017b} was applied, providing the unperturbed contrast distribution (green stars, blue diamonds and red triangles). Gaussian fits to the data (green, blue and red lines, corresponding to \unit[0]{V}, \unit[100.2]{V} and \unit[199.7]{V}, respectively) reveal comparable longitudinal coherence lengths with no significant variations for the different aperture voltages.}
\label{fig3}
\end{figure}

Fig.~\ref{fig2} (a) to (c) show the results for the first case. The tip voltage is increased stepwise starting at a negative tip voltage of $U_{SAT}$ = \unit[-1560]{V}. The biprism voltage $U_{BP} = \unit[0.331]{V}$ is kept constant. This leads to a field emission towards the first and second apertures that are both grounded in these measurements. As expected and revealed in the inset of Fig.~\ref{fig2} (a), the increase in signal behaves according to the Fowler-Nordheim theory, even though some signal is blocked by the apertures. In fig.~\ref{fig2} (b) the resulting fringe distances $s$ after interference are shown for increasing tip voltages. They decrease due to the shorter matter wave lengths $\lambda$ in combination with a lower quadrupole magnification at the higher electron energies. The fringe distances vary between $s= \unit[2.59]{mm}$ ($\lambda = \unit[31.1]{pm}$) and $s = \unit[1.89]{mm}$ ($\lambda = \unit[28.9]{pm}$) in fig.~\ref{fig2} (b). Fig.~\ref{fig2} (c) presents the interference contrast given after a second-order correlation analysis that reduces dephasing from the environment (noise from the electricity network, vibrations, slow fringe drifts etc.) \cite{Rembold2014,Guenther2015,Rembold2017,Rembold2017b}. 

The results of our study for the second case are presented in fig.~\ref{fig2} (d) to (f). Here, the tip and the biprism are set on a fixed potential of $U_{SAT} = \unit[-1600]{V}$ and $U_{BP} = \unit[0.331]{V}$ resulting in an electron matter wave length of \unit[30.7]{pm}. The voltage of the first counter electrode is varied between $U_c = \unit[-119.7]{V}$ and \unit[199.7]{V}. The last value corresponds nearly to the simulation in Fig.~\ref{fig4} (a). The second electrode is again grounded. The signal on the detector increases by a factor of $\sim 33$. A linear behavior in a Fowler-Nordheim representation of the measured signal versus the potential difference between the SAT and the counter electrode can also be observed. However, the de Broglie wavelengths and particle velocities do not vary significantly, as it is expected according to our simulations and as it can be deduced from the constant fringe distance in fig.~\ref{fig2} (e). Furthermore, the interference contrast of around \unit[50]{\%} in fig.~\ref{fig2} (f) does not change significantly, indicating constant transversal and longitudinal coherences. To perform a separate test of a possible variation of the transversal coherence length, the amount of visible fringes in the interference pattern were determined at the maximal and minimal first aperture voltage. The corresponding correlation corrected \cite{Rembold2014,Guenther2015,Rembold2017,Rembold2017b} interference pattern are illustrated in fig.~\ref{figIntpat}. The amount of counted fringes, revealing the field of coherent superposition, does not change throughout these measurements, verifying a constant transversal coherent illumination of the biprism fiber.

A possible change in longitudinal coherence for different voltages on the first aperture can be tested with the Wien filter \cite{Nicklaus1993}. Thereby, the measurements were conducted in the ``matched mode'' where the action of the electric field $\mathbf{E}$ from the Wien filter condensator cancels the one of the magnetic field $\mathbf{B}$ from the Wien coils along the optical beam axis. This ``Wien condition'' is fulfilled for: $e \, \mathbf{E} + e \, ( \mathbf{v} \times \mathbf{B} ) = 0$ with the particle velocity $\mathbf{v}$. By measuring the contrast change for different values of $\mathbf{E}$ and $\mathbf{B}$, the longitudinal coherence length can be dertermined  \cite{Nicklaus1993}. The tip and biprism voltages were kept constant at the same values as above ($U_{SAT} = \unit[-1600]{V}$, $U_{BP} = \unit[0.331]{V}$). Starting from \unit[+72]{V}, the Wien condensator voltage was stepwise decreased to \unit[-99]{V}. At each step the current in the Wien coils was increased until the fringe pattern shifts back to the original position to assure the ``matched mode''. Then a signal of several $10^5$ counts was recorded for the three different aperture voltages $U_c = \unit[0]{V}$, \unit[100.2]{V} and \unit[199.7]{V}, while only minimal phase shifts were noted from these changes of the settings. Subsequently, the average intensity along the fringe-direction was determined within a section of the spatial interference pattern. The resulting distribution was fitted with the model function \mbox{$I(x)=I_0\cdot\left(1+C\cdot\cos(\frac{2\pi x}{s}+\phi_0)\right)\cdot {\rm sinc}^2(\frac{2\pi x}{s_1}+\phi_1)$} according to a method described elsewhere \cite{Pooch2017}. Thereby, $C$ is the interference contrast and $s$ the fringe distance. The phases $\phi_0$, $\phi_1$, the average intensity $I_0$ and the width of the interference pattern $s_1$ are additional fitting parameters. The resulting contrast distributions are shown in fig.~\ref{fig3}. The data indicates that the contrast is significantly reduced by electromagnetic dephasing from the electricity network. Due to different count rates, the signal integration times were significantly longer for an aperture voltage of \unit[0]{V}, leading to a stronger dephasing compared to \unit[100.2]{V} and \unit[199.7]{V}. This causes a higher contrast loss. In fact, there is no contrast determinable at the \unit[0]{V} aperture setting for Wien filter voltages higher than $\sim$~\unit[30]{V}. For that reason, it is not possible to determine if the longitudinal coherence is preserved for different aperture voltages by the spatial interferences only. It was necessary to reveal and correct the dephasing by a second-order correlation analysis that includes the spatial and temporal differences of the electron events at the delay line detector as described in detail elsewhere \cite{Rembold2014,Guenther2015,Rembold2017,Rembold2017b}. The resulting unperturbed contrast data is also plotted in fig.~\ref{fig3}, revealing contrast rates up to \unit[71.6]{\%}. Thereby, a dephasing amplitude of $\sim$~\unit[0.4]{$\pi$} was determined. Three Gauss fits were applied to each data set. According to \cite{Nicklaus1993} the coherence length $l_c$ equals the longitudinal shift of the separated partial wave packages between two points where the contrast vanishes. This was defined to be the case when the contrast drops to \unit[10]{\%} of its maximum value. The necessary voltage $U_{cl}$ for this shift is connected to the width of the Gaussian fit $\sigma$ by $U_{cl}=\sqrt{2\cdot \ln{10}}\cdot \sigma$. For a given Wien filter condensator voltage $U_{WF}$ the shift of the wave packets is calculated by $\Delta y = \frac{L}{2D}\frac{\Delta x}{U_{SAT}} \cdot U_{WF}$, where $L$ is the length of the Wien filter condensator plates and $D$ the distance between them. $\Delta x$ denotes the distance between the separated beam paths at the center of the Wien filter \cite{Nicklaus1993}. It can be determined by $\Delta x = \Theta \cdot d_{WF-QP}$, with $d_{WF-QP}$ being the distance between the Wien filter and the quadrupole and the superposition angle $\Theta$ which can be calculated by the applied voltages $U_{SAT}$ and $U_{BP}$  \cite{Lenz1984,Pooch2017}. The resulting data reveal longitudinal coherence lengths of \unit[$82\pm9$]{nm} for \unit[$0$]{V} on the first aperture, \unit[$93\pm 10$]{nm} for \unit[$100.2$]{V} and \unit[$82 \pm 8 $]{nm} for \unit[$199.7 $]{V}. The energy widths of the emitted beams can be determined by $\Delta E = \frac{2 U_{SAT}\lambda}{\pi l_c}$. This leads to energy widths of \unit[$377 \pm 40$]{meV}, \unit[$334 \pm 37$]{meV} and \unit[$377 \pm 35$]{meV}, respectively. Our results are in good agreement with the literature value for the energy spread of SAT field emitters of \unit[0.4]{eV} \cite{Rokuta2008}. The consistency within the error bars of the longitudinal coherence lengths for different aperture voltages verifies our conclusion that the coherent beam properties are not affected by our method of intensity enhancement.

Several applications in microscopy, interferometry or sensor technology require slow coherent electrons \cite{Chang2015,Latychevskaia2017,Longchamp2015}. Our method can generate such matter waves with energies that are significantly lower than for typical field emission tips. To test the limits of the technique in our setup, we reduced the tip voltage to the values simulated in fig.~\ref{fig4} (c) and (d). Thereby, the tip voltage was set to \unit[-600]{V} were no emission is observed with a grounded counter electrode, since it is significantly lower than the minimal extraction voltage of the SAT. However, in combination with an aperture voltage of \unit[1378]{V} a reasonable count rate of \unit[$1138 \pm 2$]{Hz} after magnification was detected. The energy of the particles corresponds to a de Broglie wave length of \unit[50]{pm}. The slower the electrons are, the more susceptible they are for dephasing by external oscillations. This can also be observed in the deduced interference contrast of \unit[$19.8 \pm 1$]{\%} from the spatial integrated image compared to the determined contrast after correlation analysis of \unit[$37.7 \pm 3$]{\%}. For the same reason a larger dephasing amplitude of $\sim 0.435 \, \pi$ was determined. The measured pattern periodicity on the detector was \unit[$9.0 \pm 0.3$]{mm}. From a comparison with the theoretical value at the entrance of the quadrupole before magnification of \unit[928]{nm}, we deduce a magnification factor of $9730 \pm 290$.

\section{Conclusion}

We demonstrated in an electron biprism matter wave interferometer that the coherent properties of a beam are not affected by accelerating and decelerating the electrons or the associated lens effects. As a result, it was possible to increase the signal by a factor of $33$ at a certain matter wavelength while remaining full transversal and longitudinal coherence. It was realized by a simple design known from electron microscopy, installing a field emission tip in combination with two counter electrodes. Thereby, a single atom tip is set on a voltage well below its minimal extraction voltage. Field emission is initiated by the application of a positive voltage on a counter electrode. The second electrode is grounded and decelerates the electrons to the energy corresponding to the low tip potential. This is also verified in particle beam simulations. By measurement of the interference pattern periodicity, amount of fringes and interference contrast, it could be demonstrated that the velocity and transversal coherence of the electrons are kept constant with increasing first aperture voltage and signal intensity. Additionally, it was also possible to determine the energy width of the single atom tip emitter to be \unit[$377 \pm 40$]{meV} for \unit[$0$]{V} on the first aperture. It corresponds to a longitudinal coherence length of \unit[$82\pm9$]{nm} and does not change significantly for different counter aperture voltages. The experiment also showed that slow electrons are susceptible to external dephasing perturbations. For that reason, it was required to remove the significant dephasing from the \unit[50]{Hz} electricity network by a correlation analysis \cite{Rembold2017,Guenther2015,Rembold2014,Rembold2017b}.

The method also enabled the generation of slow coherent electrons with energies significantly lower than the ones corresponding to the minimum extraction voltage of the tip. We demonstrated this by interfering electrons with \unit[600]{eV} and a counter electrode voltage of \unit[1378]{V}, still revealing a large contrast of \unit[$37.7 \pm 3$]{\%}. Our technique is of relevance in all applications where an intense beam of tunaberable, slow and coherent electrons is required such as for microwave chip-based designs \cite{Hammer2015}, electron diffraction microscopy \cite{Chang2015,Latychevskaia2017,Longchamp2015}, sensitive sensors for inertial forces \cite{Hasselbach1993b}, vibrational \cite{Rembold2017} or electromagnetic \cite{Guenther2015} dephasing and decoherence studies \cite{Sonnentag2007,Machnikowski2006,Scheel2012,Anglin1997}.

\section*{Acknowledgements}

This work was supported by the Deutsche Forschungsgemeinschaft through the research grant STI 615/3-1 and the Vector Stiftung. A.S.~acknowledges support from the Bridging Fund of the University of T\"{u}bingen and A.R.~from the Evangelisches Studienwerk e.V.~Villigst.

\end{document}